\documentclass[a4paper,11pt]{article}

\usepackage{jcappub} 
\usepackage{subfig}
\usepackage{makecell}
\newcommand*\mean[1]{\left< #1 \right>}

\title{\boldmath A new model-independent approach for finding the arrival direction of an extensive air shower}

\author{H. Hedayati Kh.}

\affiliation{Department of Physics, K.N Toosi University of Technology, P.O. Box 15875-4416, Tehran, Iran}

\emailAdd{hedayati@kntu.ac.ir}

\abstract{A new accurate method for reconstructing the arrival direction of an extensive air shower (EAS) is described. Compared to existing methods, it is not subject to minimization of a function and, therefore, is fast and stable. This method also does not need to know detailed curvature or thickness structure of an EAS. It can have angular resolution of about 1 degree for a typical surface array in central regions. Also, it has better angular resolution than other methods in the marginal area of arrays.}

\begin{document}
\maketitle
\flushbottom

\section{Introduction}
\label{sec:intro}
It would not be exaggerated if we said that the most important property of an EAS is arrival direction. As well, the first step of reconstructing an EAS is estimation of arrival direction. EAS arrival direction is fundamental for reconstructing the core location and, more importantly, for determination of its energy. On the other hand, arrival direction mis-estimation results in systematic error of other reconstructed parameters of an EAS.\\
The most common method for finding the arrival direction of an EAS is a fit of recorded arrival times, $t^{rec}_i$s, to the expected arrival times, $t^{exp}_i$s, which is performed by minimization of the following equation:
\begin{equation}
\label{eq:CHI2}
\chi^2=\sum_{i=1}^N{w_i(t^{exp}_i-t^{rec}_i)^2}
\end{equation}
where $N$ is the number of triggered detectors (TD) of the array during an EAS event and $w_i$ is the weight which is assigned to the $i$th TD. Usually $t^{exp}$ is a plane, a cone with a fixed cone slope, a cone with a variable cone slope that is taken as a fit parameter, or a plane with curvature correction.\\
The simplest functional form of $t^{exp}$ is a plane wave front with light speed (Plane front approximation (PFA)). This plane is represented by the following equation:
\begin{equation}
\label{eq:PFA1}
\hat{\textbf{n}}\cdot(\textbf{r}-\textbf{r}_0)=c(t-t_0)
\end{equation}
where $\hat{\textbf{n}}=(n_x,n_y,n_z)$ is a unitary vector in the direction of axis of EAS, $\textbf{r}_0$ is the position of an arbitrary point on the plane, and $t_0$ is the time of arrival of EASs forward front to this point. We should only find 3 independent constants (e.g. $n_x$, $n_y$ and $t_0$) which can be seen better, if we rewrite equation \eqref{eq:PFA1} as follows:
\begin{equation}
\label{eq:PFA2}
n_xx+n_yy+n_zz=c(t-t_0)
\end{equation}
where $\textbf{r}_0$ is replaced with the coordinates of origin, $(0,0,0)$. Now, the $\chi^2$ function can be written as:
\begin{equation}
\label{eq:chi2Plane}
\chi^2=\sum_{i=1}^N{w_i(n_xx_i+n_yy_i+n_zz_i-c(t_i-t_0))^2}
\end{equation}
under the constraint of $n_x^2+n_y^2+n_z^2=1$.\\
Sometimes, all weights are taken as $w_i=1$. In these situations, summation does not often include all TDs. For example, summation is performed among a few TDs around the one recording the largest number of particles \citep{aglietta1993uhe}.\\
In some other cases, the thickness of EAS front is considered and the weights are taken as $w_i=1/\sigma_i^2$, where $\sigma_i$ is the thickness of EAS in the location of $i$th detector \citep{yoshida1995cosmic,alexandreas1992cygnus}. \cite{linsley1986thickness} established empirically that:\begin{equation}
\label{eq:sigma}
\sigma_i=1.6(\frac{r_i}{30}+1)^{1.65}\quad\text{ns}
\end{equation}
with the $r_i$, distance of $i$th detector to the core location measured in meters. When a detector detects more than one particle, the above equation should be divided by $\sqrt{n_i}$, where $n_i$ is the number of detected particles in $i$th detector.\\
Some authors prefer to consider the front of an EAS as a cone with a fixed cone slope \citep{merck1996methods}. Assuming a conical front, the equation \eqref{eq:chi2Plane} changes as follows:
\begin{equation}
\label{eq:chi2cone}
\chi^2=\sum_{i=1}^N{w_i(n_xx_i+n_yy_i+n_zz_i+s_{cone}\rho_i-c(t_i-t_0))^2}
\end{equation}
where $s_{cone}$ is the EAS cone slope and $\rho_i$ is the transverse distance of $i$th detector from the EAS axis.\\
Another possible treatment is to take the cone slope a function of EAS' other properties (e.g. a function of zenith angle \citep{acharya1993angular}). In these circumstances, at first, the EAS' parameters should be found with an initial crude estimation (e.g. arrival direction with a PFA with all $w_i$ taken as 1). A further option is to take the slope of cone as an additional fit parameter (e.g. \cite{mayer1993fast}).\\
All of the above methods of reconstructing the arrival direction of an EAS have in common minimization of a multivariable function. With the exception of the special case of a simple PFA where all $w_i$ are taken constant and whose minimization can be done analytically, all other techniques need a numerical minimization which is time-consuming and does not have a unique solution. Also, numerical minimization require a first guess for the desired parameters and because of the inherent complexity of minimization methods for a multivariable function, may not converge to a solution. Also, the same methods are partly dependent on the precision of the predicted shape of EAS front curvature or its thickness, and so are model-dependent.\\
Some of the algorithms used in the literature are model-independent and also do not make use of a minimization procedure, but are restricted to a specific array or a specific category of arrays \citep{klages1997kascade}. As an example, \cite{mayer1993fast} developed a fast gradient method which does not need a minimization procedure, but can only be used for large EASs detected in a square network array.\\
In what follows, we introduce a new arrival direction reconstruction algorithm which does not need a numerical optimization procedure and therefore is fast and stable in comparison to the above-mentioned common algorithms. This method is general, in the sense that is not restricted to a special category of surface arrays. It is also relatively accurate. This method is based on a recently introduced method for reconstructing the core location of an EAS, named SIMEFIC II \citep{hedayati2015statistical}.\section{Air Shower Simulations}\label{SimSec}
The algorithm which is presented in the next section has been developed and tested for an assumed array whose detectors and layout have been described in detail in \citep{hedayati2011new}. In short, it is a symmetric square array with $21\times21$ detectors and an array constant of 10 m. The length of each side of the array is 200 m.\\
400,000 CORSIKA version 7.4 \cite{heck1998corsika} simulated EASs whose specifications are summarized in Table ~\ref{CORSIKADef} have been used. In all simulations, EASs' true core locations move on the diagonal line of the array $(i,i)$, from the center of the array, $(0,0)$, to its corner, $(100\:\textrm{m},100\:\textrm{m})$, by steps of $(1\:\textrm{m},1\:\textrm{m})$.\\
In order to estimate the error which has been occurred in the arrival direction reconstruction of an EAS, we find the angular distance of the true arrival direction provided by CORSIKA, $(\theta_t, \phi_t)$, from the reconstructed arrival direction, $(\theta_r, \phi_r)$:
\begin{equation}
\label{eq:gamma}
\gamma=\cos^{-1}(\cos(\theta_t)\cos(\theta_r)+\sin(\theta_t)
\sin(\theta_r)\cos(\phi_t-\phi_r))
\end{equation}
\begin{table*}[h!]
\centering
\begin{tabular}{| l | c |} 
 \hline
 Specification & Value\\ [0.5ex] 
 \hline\hline
 geographical longitude & 51 E  \\ 
 geographical latitude & 35 N  \\
 altitude & 1200 m  \\
 earth magnetic field ($B_x$) & $28.1 \,\mu$T  \\
 earth magnetic field ($B_z$) & $38.4 \,\mu$T \\
 low energy hadronic model & Fluka 2011.2b \cite{ferrari2005fluka}  \\
 high energy hadronic model & QGSJETII-04 \cite{ostapchenko2011monte}  \\ [1ex] 
 \hline
\end{tabular}
\caption{EASs' specifications. The primary particle of 90\% of the showers are protons and the remaining primary particles are alphas. Other specifications are CORSIKA default values.}
\label{CORSIKADef}
\end{table*}
\section{Arrival Direction Reconstruction Algorithm}
As already mentioned, \cite{mayer1993fast} gradient method is an arrival direction reconstruction technique which does not need any optimization procedure and so is stable and fast. In this method, for every grid point of the array (position of all detectors of the array), an arrival time gradient is calculated which depends to the time of those detectors which are nearest to that grid point along two perpendicular directions ($x$ and $y$ directions). Each grid point which does not have a TD in one of its four main directions, is discarded from calculations. Then a weight which depends on pulse heights and distances of the nearest chosen detectors from others along each direction to that grid point is assigned. At last, a weighted averaging is done (for $x$ and $y$ directions) whose results are directly related to the direction cosines of the EAS.\\
One of the most striking ideas of the Mayer gradient method is that the average of gradients can compensate for the conical shape of EAS front around the axis of EAS. This procedure also reduces that contribution of arrival direction error which is the result of stochastic fluctuation of arrival times.
However, it has two main disadvantages: First, It is only applicable to square grid arrays (whose most detectors have some other neighbouring detectors along two perpendicular directions). Second, for each grid point, it is not clear why we do not use arrival time information of the same grid point (if the detector of that grid point is triggered).\\
In what follows, a new technique is introduced which has the benefits of the Mayer gradient method, but does not have its disadvantages. Instead of finding the arrival time gradient in each grid point, we find a unit vector (direction vector (DV)) in the approximate arrival direction of the EAS for each TD. Then, a weighted averaging will be done among all of these DVs. Same as Mayer gradient method, averaging can reduce the systematic error in arrival direction due to the conical shape of an EAS and the stochastic error which is the result of fluctuations in arrival times.
\subsection{Calculation of DVs}
For each TD, the DV can be found using its arrival time information and the arrival time information of a few other detectors of the array (see below for the selection rule of other detectors). This task can be accomplished with a simple PFA whose equations for a horizontal flat detector array are explicitly represented here for reference.\\
Assume we want to find DV for $m$ TDs of a horizontal flat detector array. In this case, the equation \ref{eq:chi2Plane} is changed as follows:
\begin{equation}
\label{eq:chi2DV}
\chi^2=\sum_{i=1}^m{(n_xx_i+n_yy_i-c(t_i-t_0))^2}
\end{equation}
where we are looking for $n_x$ and $n_y$. Since DV is a unit vector, the third component $n_z=+\sqrt{1-n_x^2+n_y^2}$ (apart from very rare upward EASs, $n_z$ is always positive). For finding $n_x$ and $n_y$, we should solve the following system of equations simultaneously:
\begin{equation}
\label{chi2Deriv}
\frac{\partial\chi^2}{\partial n_x}=0,\;\;\;\;\;\; \frac{\partial\chi^2}{\partial n_y}=0,\;\;\;\;\;\; \frac{\partial\chi^2}{\partial t_0}=0,
\end{equation}
Solutions of the above system of equations are:
\begin{equation}
\begin{aligned}
n_x=\frac{\mean{xy}(\mean{yt}-\mean{y}\mean{t})
+\mean{x}(\mean{y^2}\mean{t}-\mean{y}
\mean{yt})+\mean{xt}(\mean{y}^2-\mean{y^2})}
{\mean{x^2}\mean{y}^2+\mean{x}^2\mean{y^2}-2\mean{x}\mean{y}
\mean{xy}+\mean{xy}^2
-\mean{x^2}\mean{y^2}}c,\\
n_y=\frac{\mean{xy}(\mean{xt}-\mean{x}\mean{t})+\mean{y}
(\mean{x^2}\mean{t}-\mean{x}\mean{xt})+\mean{yt}(\mean{x}^2-\mean{x^2})}{\mean{x^2}\mean{y}^2+\mean{x}^2\mean{y^2}-2\mean{x}\mean{y}
\mean{xy}+\mean{xy}^2
-\mean{x^2}\mean{y^2}}c,
\end{aligned}
\end{equation}
where:
\begin{align*}
&\mean{x}=\frac{1}{n}\sum_{i=1}^{m}{x_i},
&&\mean{x^2}=\frac{1}{n}\sum_{i=1}^{m}{x_i^2},
&&\mean{xt}=\frac{1}{n}\sum_{i=1}^{m}{x_it_i},&
\end{align*}
and so on.\\
On some exceptional occasions the $n_z^2=1-n_x^2+n_y^2<0$, so we must put such data aside. The reason for this incident is that a plane with a speed of light cannot be fit to the time data of detectors.
\subsection{Weighted mean of DVs}
Here, two questions arise: how should we choose some other TDs around a TD for finding DV of the same TD, and how can we assign a weight to this DV?\\
Near core location, EAS front is flatter, and random arrival time fluctuation is smaller than other regions. So, the arrival time information of a TD which is closer to the core location is more reliable than that of detectors which are far from the core location. Two other decisive factors for reliability of a detector time information in comparison with others during an event are the high number of detected particles in that detector and its short relative distance from other detectors with high density of detected particles.
All above-mentioned criteria are included in the weights provided by SIMEFIC II for TDs. SIMEFIC II method which can reconstruct the core location of an EAS with a good precision, assigns a weight to each TD of the array. These weights are defined as $w_i=n_in_j/d_{3ij}$ (for further information see \citep{hedayati2011new} and \citep{hedayati2015statistical}). Based on these weights, the following algorithm is proposed:
\begin{enumerate}
\item Assume we have $N$ TDs during an EAS event. Using a PFA (all $w_i$s taken as 1), we find the first approximation of the arrival direction of the EAS $(\theta^\prime,\phi^\prime)$. Then, using $(\theta^\prime,\phi^\prime)$ as inputs of SIMEFIC II, we find SIMEFIC II weights for each TD of the array.
\item We sort TDs according to their SIMEFIC II weights (from now on $w_i$s are SIMEFIC II weights), from the highest one ($w_1$) to the lowest one ($w_N$).
\item Using the first $m$ highest weighted detectors’ time information, we find a DV and assign it to the $1$st highest weighted detector with the weight $w_1$. Then we eliminate the $1$st detector from the list of detectors.
\item For each remaining $N-1$ TDs, we repeat the last step one by one. At last we have $N-m$ DVs and their related weights.
\item After finding DV for each TD of the array, the reconstructed arrival direction of the EAS can be found from:
\begin{equation}
\begin{aligned}
\sin\theta_r\cos\phi_r&=\frac{\sum_{i=1}^{N-m+1}{w_in_{xi}}}{\sum_{i=1}^{N-m+1}{w_i}},\\
\sin\theta_r\sin\phi_r&=\frac{\sum_{i=1}^{N-m+1}{w_in_{yi}}}{\sum_{i=1}^{N-m+1}{w_i}},\\
\cos\theta_r&=\frac{\sum_{i=1}^{N-m+1}{w_in_{zi}}}{\sum_{i=1}^{N-m+1}{w_i}},\\
\end{aligned}
\end{equation}
where $\hat{\textbf{n}}_i=(n_{xi},n_{yi},n_{zi})$ is DV of the $i$th TD of the array.
\end{enumerate}
Because we use SIMEFIC weights, the method is named SIMEFIC Arrival Direction (SIMAD). The last thing which should be noted is that the number of selected TDs for finding each DV, $m$, should be optimized by trial and error.
\section{Optimizing SIMAD}
\begin{figure}
\centering
\resizebox{0.6\hsize}{!}
{\includegraphics[width=\hsize]{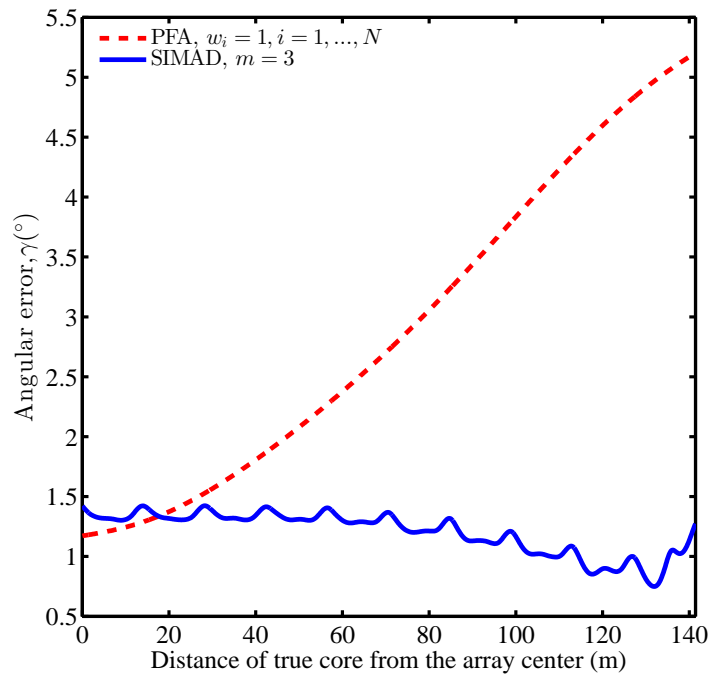}}
\caption{Comparing SIMAD ($m=3$) and PFA $w_i=1, i=1,...,N$. According to this figure, SIMAD ($m=3$) in the central part of the array has slightly worse results than PFA; but it has far better results in the marginal part of the array.
}\label{fig1}
\end{figure}

For finding the normal vector of a plane, it is necessary to have three points on it. So, the smallest possible value for $m$ is 3. Figure \ref{fig1} shows the results of SIMAD ($m=3$) and the results of a simple PFA (with $w_i=1$ for all TDs). As can be seen, it has far better results than PFA.
\subsection{Time Error}
\begin{figure}
\centering
\resizebox{0.6\hsize}{!}
{\includegraphics[width=\hsize]{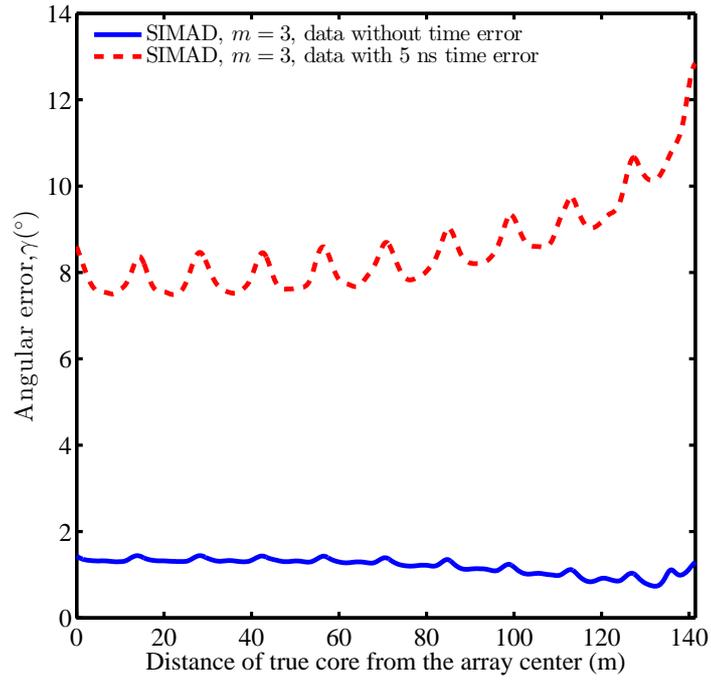}}
\caption{
Effect of $5$ns time error on the results of SIMAD ($m=3$).
\label{fig2}}
\end{figure}

\begin{figure}
\centering
\resizebox{0.6\hsize}{!}
{\includegraphics[width=\hsize]{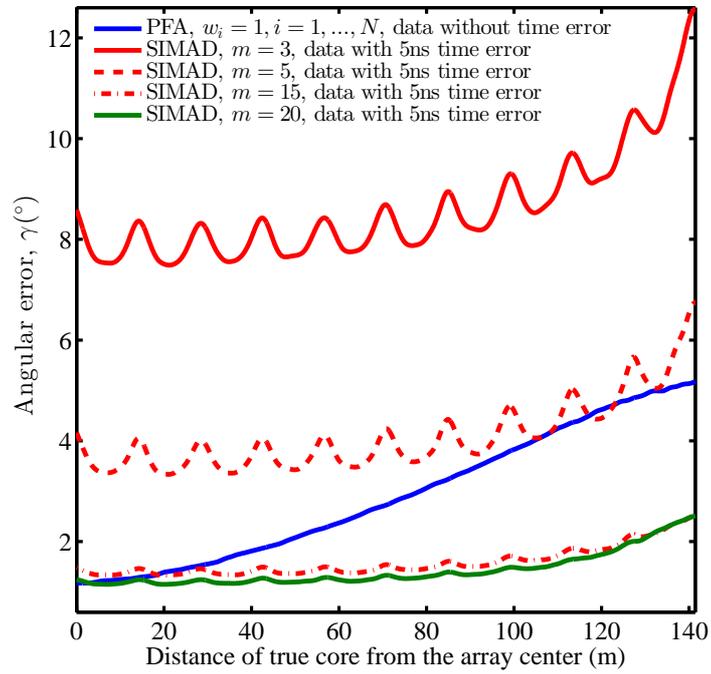}}
\caption{
The effect of increasing the amount of $m$ on accuracy of SIMAD for data with 5 ns time error.
\label{fig3}}
\end{figure}

A typical detector of a surface array with the average spacing of about $10$ m between detectors has a time resolution of $\lesssim 3$ ns (e.g. \cite{gupta2005grapes}). A sophisticated detector of such an array has a time resolution of better than 1 ns (e.g. \cite{antoni2003cosmic}). In order to check the precision of the SIMAD algorithm against detectors' limited time resolution, a normally distributed random number with an average of 0 and a standard deviation of $5$ ns was added to each particle arrival time. Figure \ref{fig2} shows the effect of this time error on the results of SIMAD ($m=3$). As can be seen, accuracy is substantially decreased by adding this time error. The reason of this behavior is not too complicated. When we find the normal vector of a plane with only 3 points on it, an uncertainty of about $1.5$ m for the positions of each point (assuming that each particle's speed is the speed of light, $5$ ns time uncertainty results in $1.5$ m of position error) leads to a large uncertainty in the direction of the plane's normal vector, especially when those points are near each other (what often happens for the near core location's detectors which have high weights).\\
In order to reduce the effect of this time error, we should increase the $m$ value. Increasing the amount of $m$ could reduce random fluctuations, because this action can play the role of an averaging with more data which reduces random error. Figure \ref{fig3} shows the results of increasing the amount of $m$ on accuracy of the SIMAD method. As you can see in this figure, with increasing the amount of $m$, the accuracy of SIMAD for data with $5$ ns time error becomes better than PFA for data without time error. Especially, for $m=20$, accuracy of SIMAD is the same as that of PFA in the central region, and is far better than PFA in peripheral regions.
\section{Comparison of SIMAD with a more sophisticated method}
\begin{figure}
\centering
\resizebox{0.6\hsize}{!}
{\includegraphics[width=\hsize]{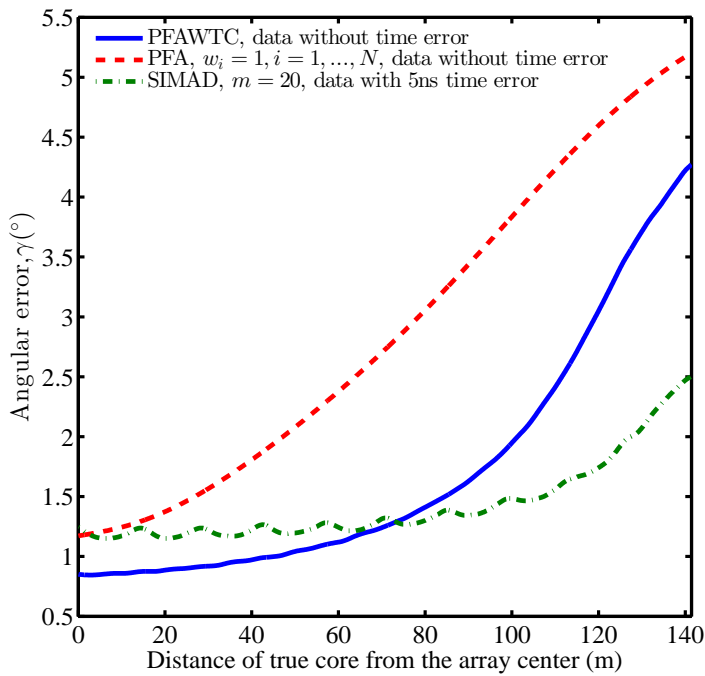}}
\caption{
Comparison of the results of SIMAD, PFA and PFAWTC.
\label{fig4}}
\end{figure}

In previous section we saw that SIMAD is superior to a simple PFA ($w_i=1, i=1,...,N$). Let us compare SIMAD with a more sophisticated method of arrival direction reconstruction than the simple PFA. A more sophisticated technique is PFA with thickness correction (PFAWTC). For the thickness of an EAS, $\sigma$, we use Linsley approximation, equation \eqref{eq:sigma}. In order to obtain better accuracy for PFAWTC, in equation \eqref{eq:sigma}, $r_i$ is taken as the distance of $i$th TD from the true axis location provided by CORSIKA, $(x_{tc},y_{tc})$:
\begin{equation}
\begin{aligned}
\Delta x=&(x_i-x_{tc})\cos\phi_t\cos\theta_t+(y_i-y_{tc})\cos\theta_t\sin\phi_t\\
\Delta y=&-(x_i-x_{tc})\sin\phi+(y_i-y_{tc})\cos\phi\\
r_i=&\sqrt{\Delta x^2+\Delta y^2}
\end{aligned}
\end{equation}
where $(x_i,y_i)$ is the location of $i$th TD and $(\theta_t,\phi_t)$ are the true angles of the arrival direction of EAS (again provided by CORSIKA). Also, data without time error are used. Figure \ref{fig4} shows the results of PFAWTC (in its ideal conditions) in comparison with PFA and SIMAD ($m=20$). For PFA, data without time error are used again. For SIMAD, data with 5 ns time error are used. As can be seen in this figure, PFAWTC is superior to PFA in all regions of the array. It also has better results than SIMAD in the central area. However, SIMAD performs better in peripheral regions.

\begin{figure}[htp!]
\begin{center}
\subfloat{
  \includegraphics[scale=0.4]{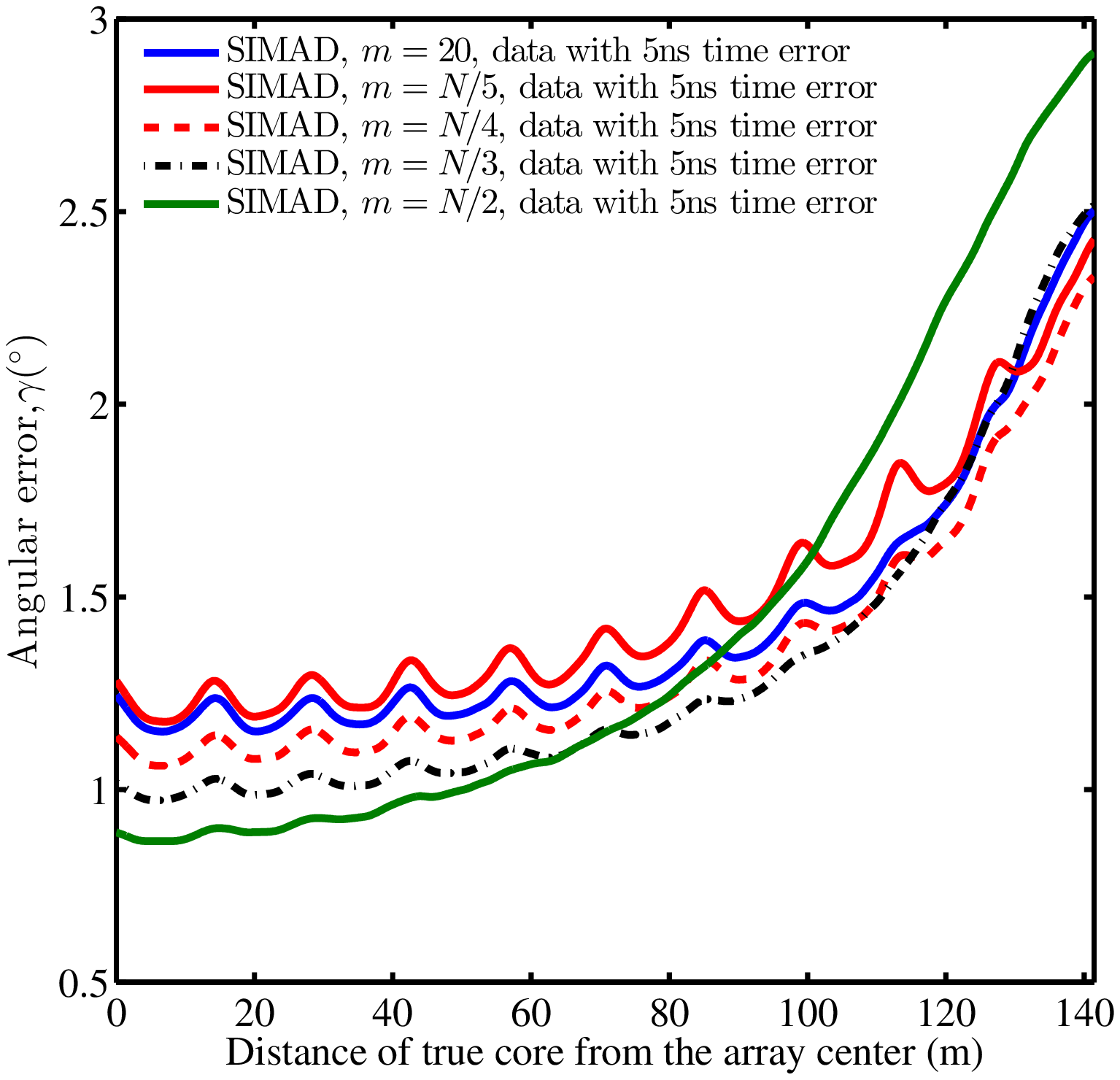}%
}\\
\subfloat{
  \includegraphics[scale=0.4]{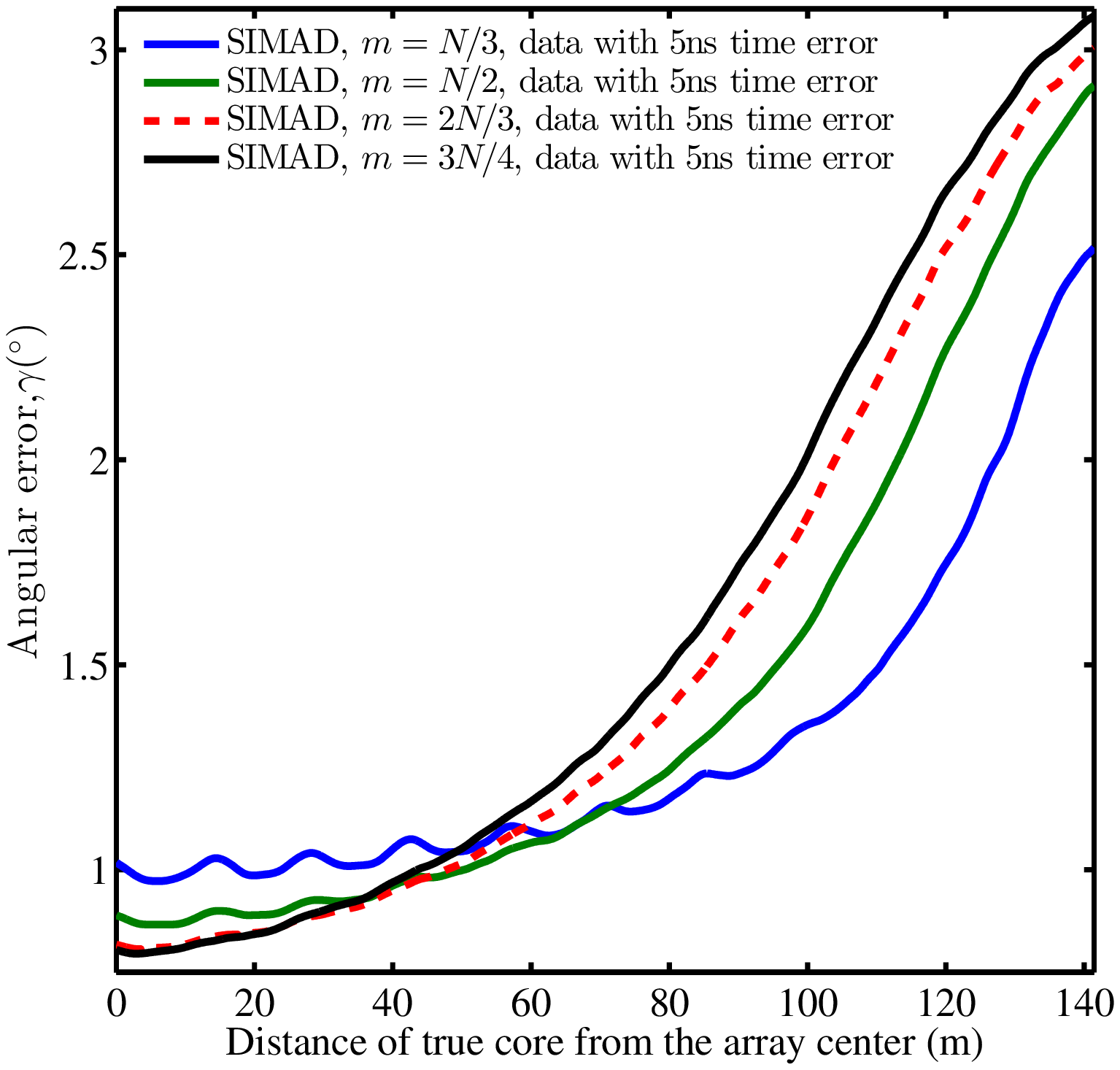}%
}\\
\subfloat{
  \includegraphics[scale=0.4]{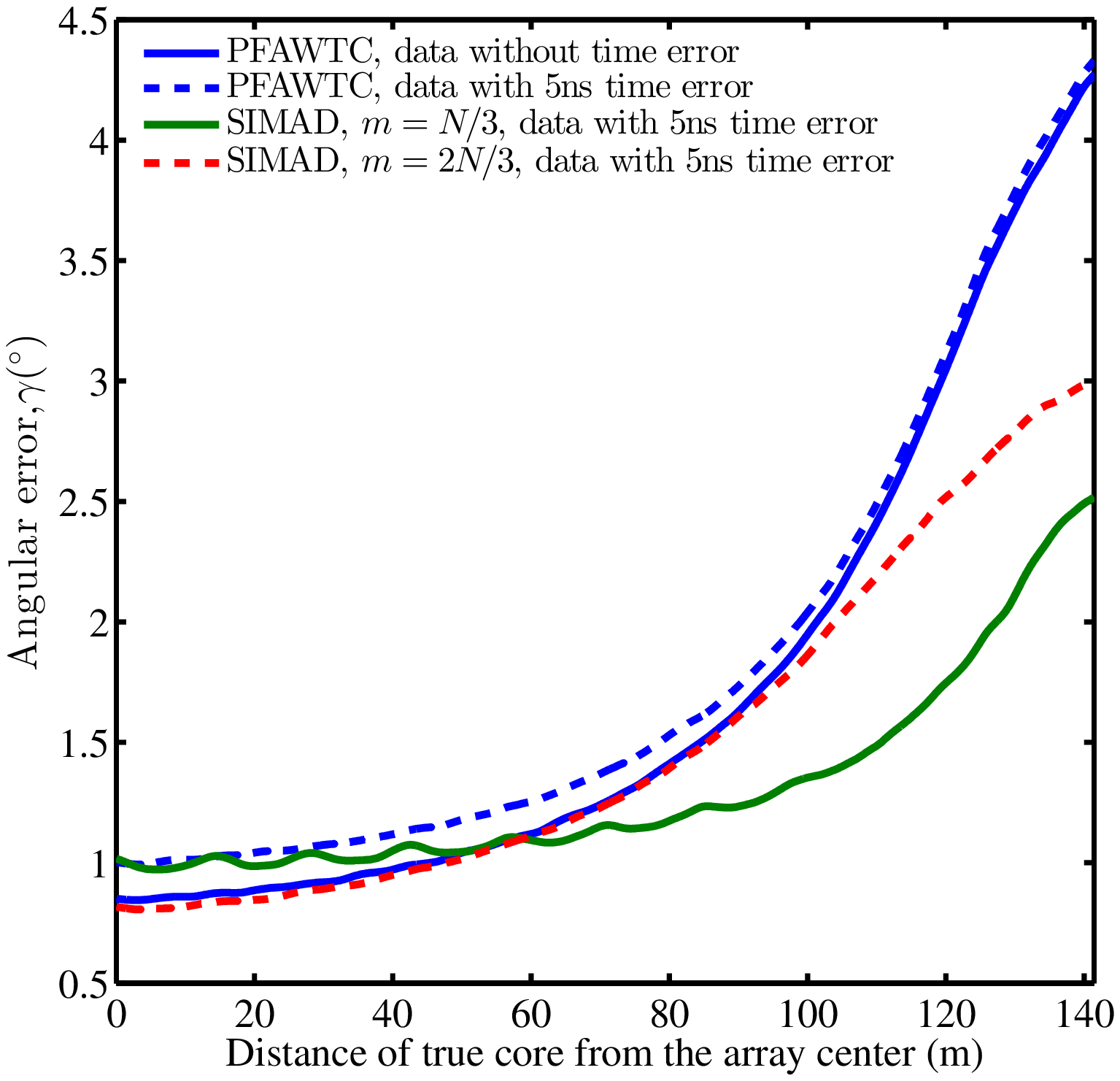}%
}
\caption{Results of SIMAD with dynamic value of $m$ proportional to the amount of $N$. In the top figure, results of SIMAD ($m=20$) and in the bottom figure, those of PFAWTC are also shown for the sake of comparison.}\label{fig5}
\end{center}
\end{figure}
\section{Further optimization}
So far, we have used the same amount of $m$ for all EASs, regardless of their sizes. It seems that for an EAS event with more TDs, the amount of $m$ should be higher in comparison with another EAS event with less TDs. Therefore, it will be better if $m$ depends on the number of TDs, $N$.\\
Figure \ref{fig5} shows the results of SIMAD for those amounts of $m$ which are proportional to the $N$. It is easy to see that increasing to some extent the amount of $m$ proportional to $N$, results in better precision of SIMAD. As is evident from the top part of the figure, results of SIMAD ($m=N/3$) are the same as or better than those of SIMAD ($N=20$) on all parts of the array. Obviously the bottom part of the figure shows that SIMAD ($m=2N/3$) has a better accuracy than PFAWTC in all parts of the array. Also, in the same part of this figure, the results of PFAWTC for data with 5 ns time error are shown. Both SIMADs ($m=N/3$ and $m=2N/3$) have better results than PFAWTC for data with time error in all regions of the array. As may be seen from the bottom part of the figure, even better results are possible if we use different amounts of $m$ in different regions of the array. Up to 100 m away from the array center, the precision of SIMAD is about $1^{\circ}$. Also, it has a better angular resolution than $2.5^{\circ}$ in the marginal part of the array.
\section{Using SIMAD for a different array}
\begin{figure}
\centering
\resizebox{0.6\hsize}{!}
{\includegraphics[width=\hsize]{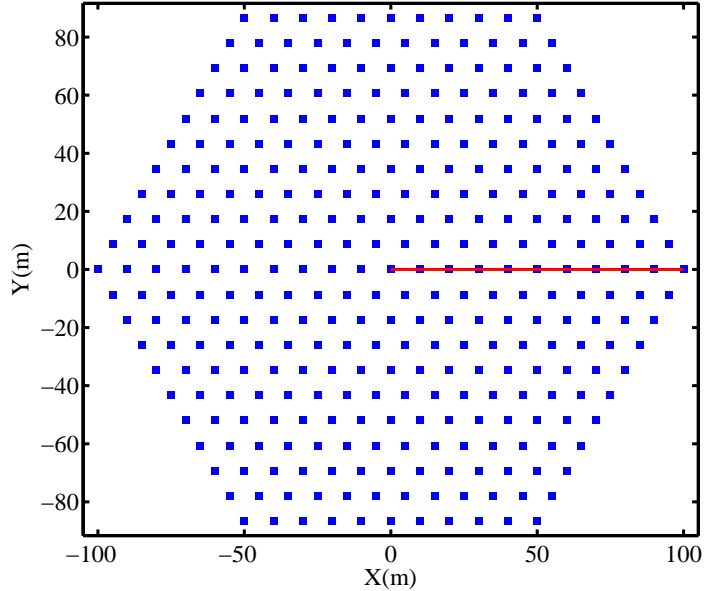}}
\caption{
Layout of a hexagonal array for the test of SIMAD method. Blue squares are the location of detectors (not to scale) whose number is 331. Other properties of the array is as the previous one. True core locations move on the red line shown in this figure.
\label{fig6}}
\end{figure}

\begin{figure}
\centering
\resizebox{0.6\hsize}{!}
{\includegraphics[width=\hsize]{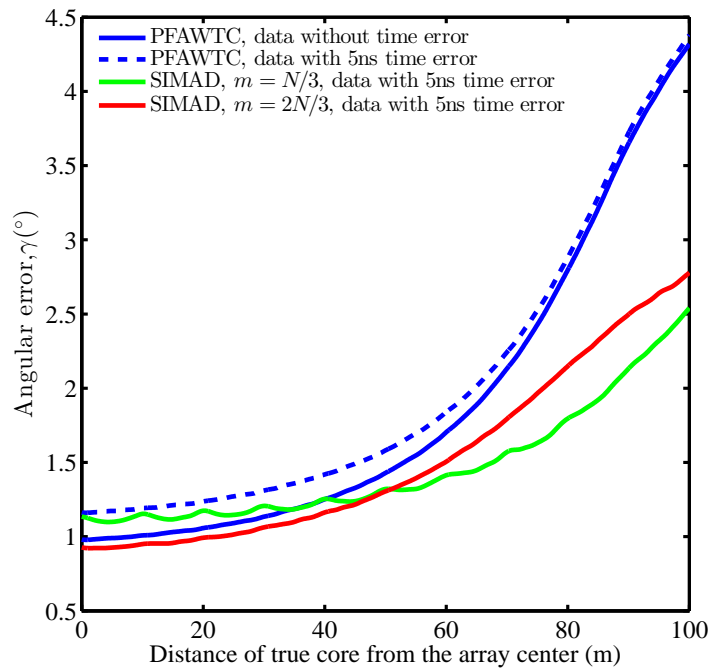}}
\caption{
Results of SIMAD and PFAWTC for the hexagonal surface array of figure \ref{fig6}.
\label{fig7}}
\end{figure}

In order to show the versatile nature of SIMAD method for different surface arrays, SIMAD method has been used for a hexagonal array layout whose detectors are arranged in a equilateral triangular network. Figure \ref{fig6} shows the layout of the array.  As before, the true core locations of EASs move on the diagonal line (red line shown in this figure).\\
Figure \ref{fig7} shows the results of  PFAWTC and also SIMAD method for this array. The same amounts of $m$ as the bottom part of figure \ref{fig5} have been used. As can be seen, SIMAD method has better results than PFAWTC for this array layout even in the central regions of the array. Although, better results can be obtained with fine tuning the values of $m$ in different regions of the array, it is clear that SIMAD even in not highly optimized form has better results than PFAWTC.\\
\section{Conclusions}
In this paper, a new technique named SIMAD for reconstructing the arrival direction of an EAS has been developed. This method does not assume anything about the shape of an EAS front or its thickness.
This technique is based on finding a local arrival direction, DV, and a special weight (provided by SIMEFIC method) for each TD of an array. The local arrival direction for a TD is found by fitting a plane to arrival times of the same TD and some other TDs around it. The weighted average direction of all TDs is a vector whose components are direction cosines of an EAS arrival direction.\\
SIMAD has a high angular resolution, especially in marginal parts of an array where other methods do not often have satisfactory precision. Also, it has at least the same accuracy of sophisticated methods in the central part of an array.\\
It should be noted that SIMAD is now in its initial version and should be optimized against different parameters and also for any other type of arrays; a few examples: weights may be not in their most optimized form; selection of some TDs for finding a local arrival direction could be improved; maybe finding a DV could be performed via some other approach; etc. Although the structure of SIMAD is general and is not dependent on a special kind of array, it should be tested and optimized for other arrays before utilizing in EASs data analysis.\\
The last point which should be insisted is that the most important advantage of SIMAD in comparison with other methods is its model-independence, so it can be used to improve other techniques of arrival direction estimation.


\begin{thebibliography}{10}
\bibitem{aglietta1993uhe}
M.~Aglietta, B.~Alessandro, P.~Antonioli, F.~Arneodo, L.~Bergamasco, A.~C.
  Fauth et~al., \emph{Uhe cosmic ray event reconstruction by the
  electromagnetic detector of eas-top}, {\emph{Nuclear Instruments and Methods
  in Physics Research Section A: Accelerators, Spectrometers, Detectors and
  Associated Equipment} {\bf 336} (1993) 310--321}.

\bibitem{yoshida1995cosmic}
S.~Yoshida, N.~Hayashida, K.~Honda, M.~Honda, S.~Imaizumi, N.~Inoue et~al.,
  \emph{The cosmic ray energy spectrum above 3$\times$ 10 18 ev measured by the
  akeno giant air shower array}, {\emph{Astroparticle Physics} {\bf 3} (1995)
  105--123}.

\bibitem{alexandreas1992cygnus}
D.~Alexandreas, R.~Allen, S.~Biller, R.~Delay, G.~Dion, X.~Lu et~al., \emph{The
  cygnus extensive air-shower experiment}, {\emph{Nuclear Instruments and
  Methods in Physics Research Section A: Accelerators, Spectrometers, Detectors
  and Associated Equipment} {\bf 311} (1992) 350--367}.

\bibitem{linsley1986thickness}
J.~Linsley, \emph{Thickness of the particle swarm in cosmic-ray air showers},
  {\emph{Journal of Physics G: Nuclear Physics} {\bf 12} (1986) 51}.

\bibitem{merck1996methods}
M.~Merck, A.~Karle, S.~Martinez, F.~Arqueros, K.~Becker, M.~Bott-Bodenhausen
  et~al., \emph{Methods to determine the angular resolution of the hegra
  extended air shower scintillator array}, {\emph{Astroparticle Physics} {\bf
  5} (1996) 379--392}.

\bibitem{acharya1993angular}
B.~Acharya, P.~Bhat, A.~John, S.~Khairatkar, B.~Nagesh, M.~Rajeev et~al.,
  \emph{Angular resolution of the kgf experiment to detect ultra high energy
  gamma-ray sources}, {\emph{Journal of Physics G: Nuclear and Particle
  Physics} {\bf 19} (1993) 1053}.

\bibitem{mayer1993fast}
H.~Mayer, \emph{A fast reconstruntion method for shower direction at large
  extended air shower arrays}, {\emph{Nuclear Instruments and Methods in
  Physics Research Section A: Accelerators, Spectrometers, Detectors and
  Associated Equipment} {\bf 330} (1993) 254--258}.

\bibitem{klages1997kascade}
H.~Klages, W.~Apel, K.~Bekk, E.~Bollmann, H.~Bozdog, I.~Brancus et~al.,
  \emph{The kascade experiment}, {\emph{Nuclear Physics B-Proceedings
  Supplements} {\bf 52} (1997) 92--102}.

\bibitem{hedayati2015statistical}
H.~Hedayati, A.~Moradi and M.~Emami, \emph{A statistical method for
  reconstructing the core location of an extensive air shower}, {\emph{The
  Astrophysical Journal} {\bf 810} (2015) 68}.

\bibitem{hedayati2011new}
H.~Hedayati, A.~Anvari, M.~Bahmanabadi, J.~Samimi and M.~K. Ghomi, \emph{A new
  method for finding core locations of extensive air showers}, {\emph{The
  Astrophysical Journal} {\bf 727} (2011) 66}.

\bibitem{heck1998corsika}
D.~Heck, G.~Schatz, J.~Knapp, T.~Thouw and J.~Capdevielle, \emph{Corsika: A
  monte carlo code to simulate extensive air showers},  tech. rep.,
  Forschungszentrum Karlsruhe {GmbH}, Karlsruhe, 1998.

\bibitem{ferrari2005fluka}
A.~Ferrari, P.~R. Sala, A.~Fasso and J.~Ranft, \emph{Fluka: A multi-particle
  transport code (program version 2005)},  tech. rep., 2005.

\bibitem{ostapchenko2011monte}
S.~Ostapchenko, \emph{Monte carlo treatment of hadronic interactions in
  enhanced pomeron scheme: Qgsjet-ii model}, {\emph{Physical Review D} {\bf 83}
  (2011) 014018}.

\bibitem{gupta2005grapes}
S.~Gupta, Y.~Aikawa, N.~Gopalakrishnan, Y.~Hayashi, N.~Ikeda, N.~Ito et~al.,
  \emph{Grapes-3—a high-density air shower array for studies on the structure
  in the cosmic-ray energy spectrum near the knee}, {\emph{Nuclear Instruments
  and Methods in Physics Research Section A: Accelerators, Spectrometers,
  Detectors and Associated Equipment} {\bf 540} (2005) 311--323}.

\bibitem{antoni2003cosmic}
T.~Antoni, W.~Apel, F.~Badea, K.~Bekk, A.~Bercuci, H.~Bl{\"u}mer et~al.,
  \emph{The cosmic-ray experiment kascade}, {\emph{Nuclear Instruments and
  Methods in Physics Research Section A: accelerators, spectrometers, detectors
  and associated equipment} {\bf 513} (2003) 490--510}.
\end{thebibliography}
\end{document}